\documentclass[a4paper,11pt]{article}
\pdfoutput=1 
\usepackage{jcappub} 

\usepackage[T1]{fontenc} 
\usepackage{longtable}
\usepackage{enumerate}
\usepackage{url}
\usepackage{array, multirow, bigdelim, makecell, booktabs}

\newcommand{\plk}{{\it Planck~}}

\title{\boldmath Constraining the average magnetic field in galaxy clusters with current and upcoming CMB surveys}

\author[a,b]{Vyoma Muralidhara}
\author[b]{and Kaustuv Basu,}

\affiliation[a]{Max-Planck-Institut f\"{u}r Astrophysik, Karl-Schwarzschild Str. 1, 85741 Garching, Germany}
\affiliation[b]{Argelander-Institut f\"{u}r Astronomie, Universit\"{a}t Bonn, D-53121 Bonn, Germany}

\emailAdd{vmura@mpa-garching.mpg.de}
\emailAdd{kbasu@uni-bonn.de}

\abstract{Galaxy clusters that host radio halos indicate the presence of population(s) of non-thermal electrons. These electrons can scatter low-energy photons of the Cosmic Microwave Background, resulting in the non-thermal Sunyaev-Zeldovich (ntSZ) effect. We measure the average ntSZ signal from 62 radio-halo hosting clusters using the \textit{Planck} multi-frequency all-sky maps. We find no direct evidence of the ntSZ signal in the \plk data. Combining the upper limits on the non-thermal electron density with the average measured synchrotron power collected from the literature, we place lower limits on the average magnetic field strength in our sample. The lower limit on the volume-averaged magnetic field is $0.01-0.24\,\mu$G, depending on the assumed power-law distribution of electron energies. We further explore the potential improvement of these constraints from the upcoming Simons Observatory and Fred Young Submillimeter Telescope (FYST) of the CCAT-prime collaboration. We find that combining these two experiments, the constraints will improve by a factor of two, which can be sufficient to rule out some power-law models.}

\begin{document}
\maketitle
\flushbottom

\section{Introduction}
\label{sec:intro}

Galaxy clusters (GCs) are the largest gravitationally bound aggregation of matter with masses up to $~10^{15}\,M_\odot$, formed at the nodes of filaments in the cosmic web. They are formed through mergers of smaller clusters and groups of galaxies, and through accretion \cite{Kravtsov2012} from the intergalactic medium surrounding filaments of galaxies (e.g., \cite{Peebles1970, Press1973, Voit2004}). The intracluster medium (ICM) is the primary reservoir of baryons within these nodes, accounting for roughly 15\% of the total cluster mass \cite{White1993, Vikhlinin2006}, and exhibiting a complex interplay between hot ionized plasma, turbulence, and an underlying extended magnetic field \cite{Sarazin1986}.
\paragraph{}

These astrophysical processes determine the observable properties of GCs and multi wavelength observations are necessary to understand the roles of the different ICM components. The dominant (and cosmologically relevant) ICM component is the bulk plasma following Maxwell-Boltzmann distributions within a range of temperatures, made visible by the thermal bremssstrahlung emission in the X-rays, or from the thermal Sunyaev-Zeldovich (tSZ) \cite{Sunyaev:1970bma, Sunyaev:1972eq} distortion in the cosmic microwave background (CMB). To a large extent, this dominant thermal component remains unaffected by the presence of the non-thermal particles and magnetic fields, although in specific, localized regions such as radio lobes, the latter can dominate the plasma dynamics \cite{Prokhorov2010, Battaglia2011, Eckert2018, Bykov2019}. However, as the sensitivity of our measurements improve, the modeling uncertainties arising from these non-thermal components will play an increasingly important role. Further, their understanding can offer new insights to probe the origin and dynamics of the large-scale structure of the Universe.
\paragraph{}

The main evidence that non-thermal electrons and magnetic fields exist in the inter-galactic space in GCs comes from different types of observations of diffuse synchrotron emission at radio wavelengths \cite{Brunetti2014, vanweeren2019}. Typically, the observed morphology of diffuse synchrotron emission can be classified into (i) cluster radio relics which are of irregular shape and trace merger shocks, (ii) radio halos which are centrally located and generally much more extended than the relics, and (iii) revived active galactic nucleus (AGN) fossil plasma sources which trace AGN plasma re-energised by various physical processes in the ICM. In this work we focus on the radio halos (RHs), as these are the only truly cluster-wide non-thermal emission whose morphologies have been shown to follow closely that of the ICM (e.g., \cite{Govoni2001, Rajpurohit2018, Botteon2020}). 
While the origin of RHs remains unclear, a general consensus has arisen behind a turbulent re-acceleration model, in which populations of seed electrons are locally re-accelerated due to turbulent states of the ICM, following the case of GC mergers (e.g., \cite{Pasini2022, Ruszkowski2023}). Despite its observational success over competing theories, the turbulent re-acceleration model suffers from uncertainty about the source and the energy distribution of the seed electrons that need to be fixed \textit{posteriori} from observational data. A direct measurement or constraints on the cluster-wide non-thermal electron spectral energy distribution (SED) is therefore a much-valued quantity.
\paragraph{}

Magnetic field strength in the diffuse ICM is measurable from the observations of the Faraday Rotation Measure (FRM) (which is inferred from observations of polarized synchrotron emission at multiple wavelengths) of the embedded or background radio sources with intrinsic polarization \cite{Heiles1976, Verschuur1979}. The synchrotron emission alone cannot be used to directly estimate the magnetic field, as it depends on the product of the non-thermal particle density and some power of the magnetic field strength. The FRM data remains sparse due to a lack of suitably positioned background sources at cosmological distances, and is also sensitive to the local environment of the polarized sources, susceptible to biases arising from the location of polarized sources, and foregrounds \cite{Johnson2020, Osinga2022}. 
In this regard, measurement of the inverse-Compton (IC) emission in combination with the synchrotron emission has been considered as the most promising way to constrain cluster-wide magnetic fields, as the former depends only on the SED of the non-thermal electrons (when the incoming radiation source is known), and helps to break the degeneracy with the magnetic field strength in the synchrotron data. The predominant case of incoming radiation is the CMB, which when scattered by the $\sim$GeV energy non-thermal electrons, results in the excess IC emission that extends to X-ray and gamma-ray regimes \cite{Sarazin1999, Wik2012, Ackermann2015, Xi2017}.

\paragraph{}

Measurement of the excess IC emission in the X-rays has been a decades-long endeavour, with mixed success \cite{Rephaeli2006, Million2008, Ota2013, Mernier2023}. The main difficulty lies in the limited sensitivity of the X-ray instruments in the hard X-ray energies, which is absolutely critical for distinguishing the IC emission component from the multi-temperature and multi-keV plasma's thermal emission \cite{Wik2012, Wik2014, Cova2019, RojasBolivar2020}. In this regard, the measurement of the same IC effect in the millimeter/submillimeter domain is now poised to make a decisive contribution, in light of the unprecedented depth of many recent and upcoming CMB sky surveys. The relevant physical phenomenon is the non-thermal Sunyaev-Zeldovich (ntSZ) effect, which, in contrast to the dominant tSZ effect, concerns the scattering of CMB photons from the non-thermal electron populations \cite{Ensslin2000, Mroczkowski:2018nrv}. There is a long history of ntSZ research in the context of GCs and radio lobes of AGN \cite{Ensslin2000, Colafrancesco2003, Colafrancesco2008, Colafrancesco2012, Acharya2020, Marchegiani2021}, although no direct detection has been made of the global ntSZ signal in GCs, apart from one measurement localized to known X-ray cavities in the ICM \cite{Abdulla2019}.
\paragraph{}

Our goal in this paper is to show that the current and upcoming CMB data are very close to making a measurement of the global IC excess,  and we place meaningful constraints on the magnetic fields in GCs from these data. Specifically, we study whether the \textit{Planck} satellite’s all-sky survey data and newer catalogs of radio halo clusters (i) can provide any constraints on the SED of the non-thermal electrons from modelling the ntSZ signal, and (ii) can potentially place constraints on the magnetic field strength by combining these constraints with the existing synchrotron flux measurements. From there, we explore the constraining power of upcoming CMB experiments such as Simons Observatory (SO) \cite{SimonsObservatory2018} and Fred Young Submillimeter Telescope (FYST) \cite{CCAT-Prime2021} on the ntSZ effect and further on the magnetic field strengths.
\paragraph{}

The rest of this paper is organized as follows. In Section \ref{sec:theory}, we formulate the modelling of the ntSZ effect and synchrotron emission in GCs, and discuss the assumptions we have made in this work. In Section \ref{sec:data}, we discuss the data and simulated microwave sky maps from which current and future constraints on the ntSZ signal, the non-thermal electron density, and magnetic field strength are measured, respectively. We discuss the methods implemented in the extraction of the SZ spectrum from \textit{Planck} data, and the fitting procedure in Section \ref{sec:fit}. We present the results in Section \ref{sec:results} which are then discussed in Section \ref{sec:summary}. We assume a flat $\Lambda$CDM cosmological model with the parameter values $\Omega_m = 0.308$ and $H_0 = 67.8\,\mathrm{km}\,\mathrm{s}^{-1}\mathrm{Mpc}^{-1}$ \cite{Planck2015cosmo} throughout this paper.  


\section{Theoretical basis}
\label{sec:theory}
This section describes the theoretical framework of our analysis. As outlined in the introduction, our method of finding the signature of the ntSZ signal or placing upper limits on non-thermal electron density (which translates to lower limits on the magnetic field strength) is based on three assumptions:
\begin{enumerate}[i)]
\item non-thermal electron pressure follows the same radial distribution as the thermal electron pressure,
\item non-thermal electrons follow one single power-law momentum distribution throughout the cluster volume, and 
\item the magnetic field energy density follows the number density of non-thermal electrons.
\end{enumerate}

These assumptions can be considered too simplistic to capture the complexity of the ICM, but they simplify the data analysis and allow us to place meaningful first constraints. Furthermore, there is some degree of theoretical and observational support for at least the first and third assumptions. Below we describe our motivation behind adopting these three criteria.

The first assumption on spatial distribution of non-thermal electrons allows us to create a 2D matched filter (Section \ref{sec:fit}) to optimally extract the cluster ntSZ signal, along with the thermal SZ (tSZ) signal, from the maps. Combined with the second assumption, this also enables us to obtain the density profile of non-thermal electrons by assuming that they have the same pseudo-temperature (Section \ref{sec:pseudoT}) throughout the emitting volume. Evidence that the non-thermal pressure density closely follows that of thermal electrons have been shown in several simulations of CR transport (\cite{Pinzke:2010st, Zandanel:2013wja}). We specifically refer to the results from \cite{Pinzke:2010st} which show that the ratio $X_\mathrm{CR} = P_\mathrm{CR}/P_\mathrm{th}$ stays approximately constant, within a narrow range, for a Coma-like disturbed cluster out to a large fraction of the virial radius. Since our sample of RH clusters consists of disturbed systems, it will be reasonable to assume that the non-thermal pressure profile thus closely follows that of the thermal pressure.
\paragraph{}
The second assumption is merely a tool for simplifying the calculations, although it can easily be relaxed for more complicated models. By assuming a uniform, global power-law distribution we ignore the effects of electron ageing, reacceleration etc., however, we do compare results for four different power-law distributions to assess the impact of this simplistic assumption.
\paragraph{}
Lastly, the third assumption of a universal magnetic field radial profile in RH clusters is not critical for our analysis, but it enables a more realistic calculation of the synchrotron power and comparison with radio data, as opposed to assuming a constant $B$ value. This energy equipartition argument leads to magnetic field strength scaling roughly to the square-root of the thermal electron density, $B(r) \propto n_\mathrm{e,th}(r)^{0.5}$. Observational evidence for such a scaling have been found by \cite{Murgia2004,Bonafede2010}
and discussed in the context of MHD simulations by \cite{Vazza2018}.

\subsection{Characteristics of the ntSZ effect}
\label{sec:ntsz}
The distortion in specific intensity due to the ntSZ effect can be written as
\begin{equation}
    \delta i(x) = (j(x) - i(x))\: I_0 \: \tau_{\mathrm{e,nth}}, \quad \tau_{\mathrm{e,nth}} = \sigma_T\int n_{\mathrm{e,nth}} dl,
    \label{eq:di1}
\end{equation}
where $x=\frac{h\nu}{k_\mathrm{B}T_{\mathrm{CMB}}}$, $I_0 = 2\frac{(k_\mathrm{B}T_{\mathrm{CMB}})^3}{(hc)^2}$ is the specific intensity of the CMB, $i(x) =\frac{x^3}{e^x-1}$ is the Planck spectrum attributed to the CMB spectrum, $\tau_{\mathrm{e,nth}}$ is the optical depth due to non-thermal electrons, and $j(x)$ is the flux scattered from other frequencies to frequency x. For a given isotropic electron momenta distribution $f_\mathrm{e}(p)$ (where p is the normalized electron momentum, $p=\frac{p_{\mathrm{phys}}}{m_\mathrm{e}c}$ and $p_{\mathrm{phys}}=\beta_\mathrm{e}\gamma_\mathrm{e}$) with normalization $\int_0^\infty f_\mathrm{e}(p)p^2\,dp = 1$, the ntSZ effect can be described as \cite{Ensslin2000}
\begin{equation}
  \delta i(x)  = \Bigg[\Bigg(\int_{p_1}^{p_2}\int_{-s_\mathrm{m}(p_1)}^{s_\mathrm{m}(p_2)}f_\mathrm{e}(p) K(e^\mathrm{s};p)\,e^\mathrm{s}\, \frac{(x/e^\mathrm{s})^3}{(e^{x/e^\mathrm{s}}-1)}\, ds\, dp \Bigg) - \frac{x^3}{e^x-1}\Bigg]I_0\tau_{\mathrm{e,nth}},
  \label{eq:di2}
\end{equation}
where $s_\mathrm{m} (p) = \mathrm{ln}\Big[\frac{1+\beta_\mathrm{e}}{1-\beta_\mathrm{e}}\Big]$ is the maximum logarithmic shift in energy with $\beta_\mathrm{e}=\frac{p}{\sqrt{1+p^2}}$ and the photon scattering kernel \cite{Ensslin2000}
\begin{equation}
\begin{split}
    K(e^\mathrm{s};p) = &-\frac{3(1-e^\mathrm{s})}{32p^6e^\mathrm{s}}\big[1+(10+8p^2+4p^4)e^\mathrm{s}+e^{2s}\big]\\
    &+\frac{3(1+e^\mathrm{s})}{8p^5}\Bigg[\frac{3+3p^2+p^4}{\sqrt{1+p^2}}-\frac{3+2p^2}{2p}(2\mathrm{arcsinh}(p)-\arrowvert s\arrowvert)\Bigg].
\end{split}
    \label{eq:ksp}
\end{equation}
The amplitude and shape of the spectrum of the ntSZ effect is dependent on the number density and the momentum distribution of the scattering non-thermal electrons. In this work, we consider power-law and broken power-law models for the scattering non-thermal electrons with different minimum and maximum momenta.

\subsubsection{Power-law distribution} 
The simplest and most commonly used distribution of non-thermal electron momenta would be a negative power-law, with fixed minimum ($p_1$) and maximum ($p_2$) momenta, and power-law index ($\alpha$). Imposing the normalization of $\int_0^\infty f_\mathrm{e}(p)p^2\,dp = 1$, this power-law is written as
\begin{equation}
    f_\mathrm{e}(p;\alpha,p_1,p_2) = A(p_1, p_2, \alpha)p^{-\alpha}, \quad \mathrm{where} \quad A(p_1, p_2, \alpha)=\frac{(\alpha -1)}{(p_1^{1-\alpha} - p_2^{1-\alpha})}.
    \label{eq:electrondist1}
\end{equation}
With the assumption that the same scattering electrons cause synchrotron radiation, $\alpha$ is related to the spectral index of synchrotron emission, $\alpha_{\mathrm{synch}}$, as $\alpha = 2\alpha_{\mathrm{synch}}+1$ \cite{Rybicki}.

This simple case can be improved by considering a broken power-law to mimic radiative energy losses at the low-energy end of the spectrum. 
For modelling the distribution of non-thermal electron momenta with a broken power-law, we fix the minimum ($p_1$), break ($p_{\mathrm{br}}$) and maximum ($p_2$) momenta, and take $\alpha_1$ and $\alpha_2$ as the indices of the flat and power-law parts of the model. This broken power-law can then be written as \cite{Colafrancesco2003}
\begin{equation}
f_\mathrm{e}(p;p_1,p_2,p_{\mathrm{br}},\alpha_1,\alpha_2)=
   C(p_1,p_2,p_{\mathrm{br}},\alpha_1,\alpha_2)
    \begin{cases}
      p^{-\alpha_1} & p_1 < p < p_{\mathrm{br}} \\
      p_{\mathrm{br}}^{-\alpha_1+\alpha_2}p^{-\alpha_2} & p_{\mathrm{br}} < p < p_2\\
      \end{cases}.
      \label{eq:electrondist2}
\end{equation}
As with the power-law model, we consider $\alpha_2 = 2\alpha_\mathrm{synch}+1$, and the normalization factor $C(p_1,p_2,p_{\mathrm{br}},\alpha_1,\alpha_2)$ arises due to the condition that $\int_0^\infty f_\mathrm{e}(p)p^2\,dp = 1$. We choose a small, non-zero power-law index for the ``flat'' part of the broken power-law model for ease of numerical integration.
\paragraph{Adopted model parameters:} 
We consider four different cases of electron momentum distribution in this paper: Two single power-law distributions with $p_1=30$ and 300, respectively (cases S1 and S2); and two broken power-law distributions with $p_\mathrm{br}$=300 and 1000, respectively (cases B1 and B2). We fix $\alpha_\mathrm{synch}=1.3$, meaning the indices of the power-laws are fixed to $\alpha = \alpha_2 = 3.6$ in Eqs. (\ref{eq:electrondist1}) and (\ref{eq:electrondist2}). Together with a dominant thermal component with ICM temperature 8 keV, whose momentum is characterized by a Maxwell-J\"{u}ttner distribution (Appendix \ref{app:rSZ}), these model parameters are used in turn to fit the match-filtered peak signal. These model parameters are summarized in Table \ref{tab:models}.

\begin{figure}[ht]
    \centering
    \includegraphics[width=1.0\textwidth]{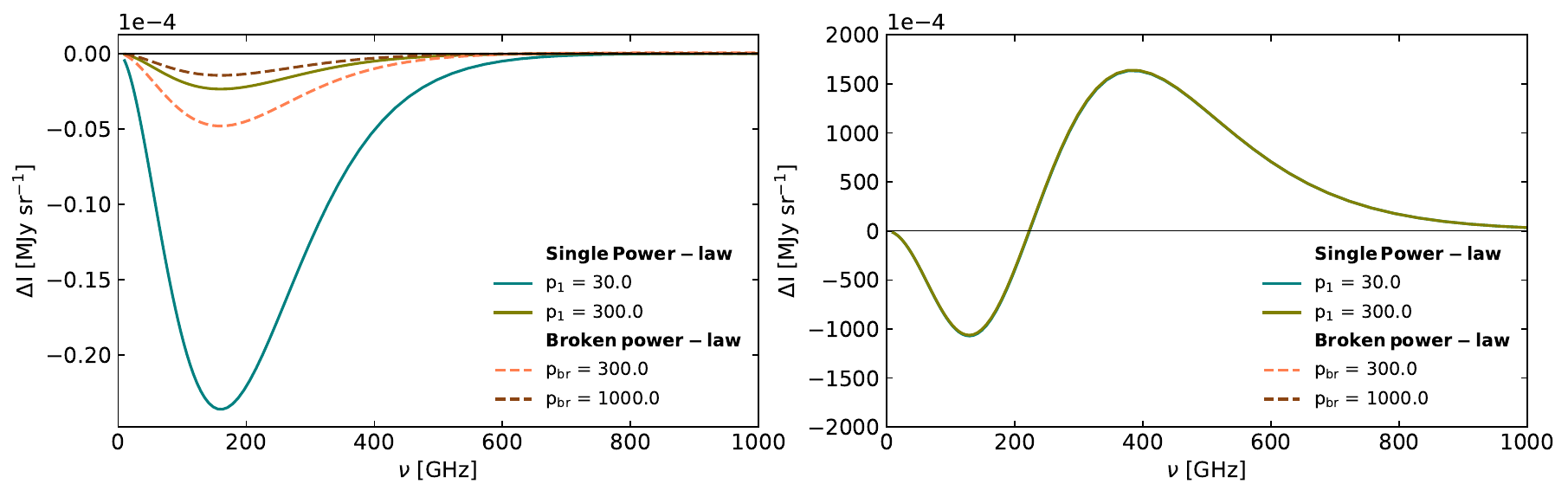} 
    \caption{\textit{Left}: Distortion in the CMB specific intensity introduced by the ntSZ effect due to \textit{single} power-law and \textit{broken} power-law non-thermal electron models with $y_\mathrm{nth}=10^{-6}$. Amongst the models considered, the amplitude of the distortion is largest for the single power-law with $\mathrm{p_{1}}=30$. \textit{Right}: The total SZ spectrum that consists of the tSZ$_\mathrm{rel}$ with $k_\mathrm{B}T_\mathrm{e}=8.0\,$keV and $y_{th}=10^{-4}$, kSZ effect with $v_\mathrm{pec}$ derived from a Gaussian distribution with $\sigma=100\, \mathrm{km\,s}^{-1}$, and ntSZ effect estimated for each of the non-thermal electron models with $y_\mathrm{nth}=10^{-6}$. The distinction from the dominant tSZ effect is not visible in this linear-scale plot.}
    \label{fig:ntSZ-spectra}
\end{figure}

\begin{table}[ht]
        \centering
    \begin{tabular}{ccc}
    \hline
     Components & Model & Parameters \\
    \hline
    &  & \\
    $\mathrm{tSZ_{rel}}$ ($k_\mathrm{B} T_\mathrm{e}$ = 8 keV)  &   \hspace{-1em}\rdelim\{{2}{*}
    S1 & $p_1 = 30$,  $p_2 = 10^5$, 
    $\alpha=3.6$ \\
    
    $+$ {\it single} power-law &   S2 & $p_1 = 300$,  $p_2 = 10^5$, $\alpha=3.6$ \\ 
    
    & & \\ 
    
    $\mathrm{tSZ_{rel}}$ ($k_\mathrm{B} T_\mathrm{e}$ = 8 keV)  &   \hspace{-1em}\rdelim\{{2}{*}
    B1 & $p_1 = 1$,  $p_\mathrm{br} = 300$, $p_2 = 10^5$,  $\alpha_1=0.05$, $\alpha_2 = 3.6$ \\
    
    $+$ {\it broken} power-law &   B2 & $p_1 = 1$,  $p_\mathrm{br} = 1000$, $p_2 = 10^5$, 
    $\alpha_1=0.05$, $\alpha_2 = 3.6$  \\    
    & & \\ 
    \hline    
       \end{tabular}
            \caption{Adopted parameters for the single and broken power-law models, along with the fixed-temperature thermal component, that are used in the spectral fitting.}
    \label{tab:models}
\end{table}
\paragraph{} 
The distortion in the CMB specific intensity introduced by the ntSZ effect, with the assumption of non-thermal electron models described in Table \ref{tab:models}, is shown in Figure \ref{fig:ntSZ-spectra}. The distortion due to the S1 model is the largest as, under our definition of the normalization of the electron momenta distributions, more non-thermal electrons are available to scatter the CMB photons. We also notice that the spectra are shallower for higher $p_1$ or $p_\mathrm{br}$ and the frequency at which the distortion is zero is shifted to higher frequencies. In the \textit{right} panel of Figure \ref{fig:ntSZ-spectra}, we also show the total SZ effect (tSZ$_\mathrm{rel}$, kSZ (described in Appendix \ref{app:ksz}) and ntSZ) wherein we see the characteristic shape of the spectrum of the SZ effect with a decrement in the specific intensity of the CMB at frequencies $<217$ GHz and an increment at frequencies $> 217$ GHz. The tSZ$_\mathrm{rel}$ is the dominant effect and thus, it is difficult to disentangle the distortions due to the other SZ effects.

\subsubsection{The zero-crossing frequency}
\begin{figure}[h]
    \centering
    \includegraphics[width=\textwidth]{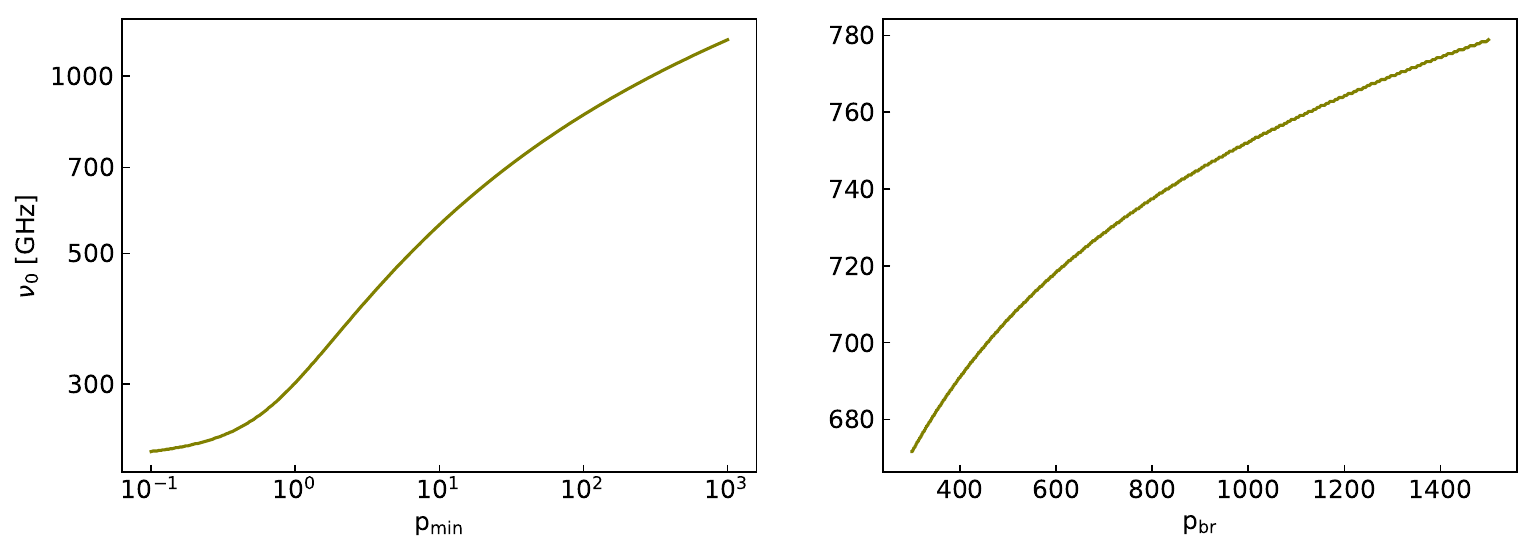}
    \caption{Frequency of zero-distortion in  specific intensity as a function of p\textsubscript{1} (left) and p\textsubscript{br} (right) which are parameters used to define the power-law and broken power-law models to compute the ntSZ effect [Eqs.  (\ref{eq:electrondist1}$-$\ref{eq:electrondist2})].}
    \label{fig:nuzero}
\end{figure}

Observations at submillimeter frequencies (roughly, above 300 GHz) are important for finding the spectral signature of the ntSZ signal. A characteristic feature of any inverse-Compton spectral distortion is the frequency at which there is no net distortion. For the ntSZ effect, this zero-crossing frequency is sensitive to the lower momentum cut-off of the electron momentum distribution, essentially the energy density of the non-thermal electrons (as shown in Figure \ref{fig:nuzero}). By measuring the zero-crossing frequency from observed spectra, one can distinguish between the energy densities of the thermal and non-thermal electron populations in the ICM. With prior information on the temperature of the population of thermal electrons, constraints on the momentum distribution of non-thermal electrons can be obtained.

\subsubsection{Pseudo-temperatures and non-thermal pressure}
\label{sec:pseudoT}
Analogous to the Comptonization parameter associated with the thermal SZ effect, we can express $\tau_{\mathrm{e,nth}}$ in terms of $y_{\mathrm{nth}}$ as \cite{Ensslin2000}
\begin{equation}
    \tau_{\mathrm{e,nth}} = \frac{m_\mathrm{e}c^2}{\langle k_\mathrm{B}\tilde{T}_\mathrm{e}\rangle}\:y_{\mathrm{nth}},
    \label{taunth}
\end{equation}
where
\begin{equation}
  y_{\mathrm{nth}}  = \frac{\sigma_T}{m_\mathrm{e}c^2} \int n_{\mathrm{e,cr}} k_\mathrm{B}\tilde{T}_\mathrm{e} dl,
  \label{ynth}
\end{equation}
is the integral of the non-thermal electron pressure along line-of-sight and
\begin{equation}
    k_\mathrm{B}\tilde{T}_\mathrm{e} = \frac{P_{\mathrm{e,nth}}}{n_{\mathrm{e,cr}}} = \int_0^\infty f_\mathrm{e}(p)\frac{1}{3}p\: v(p)\: m_\mathrm{e}c\:dp.
    \label{pseudo}
\end{equation}
Here, $k_\mathrm{B}\tilde{T}_\mathrm{e}$ is a pseudo-temperature attributed to the non-thermal electrons and $n_{\mathrm{e,cr}}$ is the normalization of the number density of non-thermal electrons. An analytical expression for Eq.\ \eqref{pseudo} which is given by \cite{Ensslin2000},
\begin{equation}
    k_\mathrm{B}\tilde{T}_\mathrm{e} = \frac{m_\mathrm{e}c^2(\alpha-1)}{6[p^{1-\alpha}]_{p_{2}}^{p_{1}}}\Bigg[\mathrm{B}_{\frac{1}{1+p^2}}\Big(\frac{\alpha-2}{2},\frac{3-\alpha}{2}\Big)\Bigg]_{p_{2}}^{p_{1}},
    \label{pseudo2}
\end{equation}
where  $\mathrm{B_x}(a, b)$ is the incomplete beta function. Rewriting Eq. (\ref{eq:di2}) in terms of $y_\mathrm{{nth}}$, we obtain
\begin{equation}
    \delta i(x) = \Bigg[\Big(\int_{p_1}^{p_2}\int_{-s_\mathrm{m}(p_1)}^{s_\mathrm{m}(p_2)}f_\mathrm{e}(p)\:K(e^\mathrm{s};p)\,e^\mathrm{s}\,ds\: dp \Big) \,-\, \frac{x^3}{e^x-1}\Bigg]\, I_0\frac{m_\mathrm{e}c^2}{\langle k_\mathrm{B}\tilde{T}_\mathrm{e}\rangle}y_\mathrm{{nth}}.
    \label{eq:di3}
\end{equation}

Since the pseudo-temperature is fixed by the choice of the power-law momentum distribution, the non-thermal electron distribution is "isothermal" in our analysis. Correspondingly, the density profile follows that of the assumed GNFW model (Section \ref{sec:gnfw}) of ICM pressure, which is then converted into a synchrotron emissivity profile using a magnetic field-strength model.

\subsection{Synchrotron emission}
\label{sec:sync}
The energy lost by an electron with an arbitrary pitch angle ($\theta$) in the presence of a magnetic field with strength $B$ is \cite{Rybicki}
\begin{equation}
P_{\mathrm{emitted}}(\nu) = \frac{\sqrt{3}e^3\,B\,\mathrm{sin}\theta}{m_\mathrm{e}\,c^2}\,x\int_x^\infty d\xi \, K_{5/3}(\xi),
    \label{eq:synch1}
\end{equation}
where
\begin{equation}
   x = \frac{\nu}{\nu_c}, \quad  \nu_c = \frac{3e\,B\,\gamma^2}{4\pi m_\mathrm{e}c}\,\mathrm{sin}\theta,
   \label{eq:nu-c}
\end{equation}
and $K_{5/3}(\xi)$ is the modified Bessel function of second kind of order 5/3. Consider a power-law distribution of electrons written as\footnote{If the distribution is assumed to be locally isotropic and independent of pitch angle, it reduces to $N(\gamma)=k\gamma^{-\alpha}$.}
\begin{equation}
    N(\gamma,\theta) = \frac{k}{4\pi}\,\gamma^{-\alpha}\,\frac{\mathrm{sin}\theta}{2}, \quad \gamma_1<\gamma<\gamma_2.
    \label{electronspec}
\end{equation}
The total synchrotron emission per unit volume for such a distribution of electron momenta is then given by
\begin{equation}
\begin{split}
    \frac{dW}{d\nu\,dt} &= \int\int P_{\mathrm{emitted}}(\nu)\,N(\gamma,\theta)\,d\gamma\,d\Omega_\theta \\
    &= \frac{\sqrt{3}k\,e^3B}{8\pi m_\mathrm{e}c^2}\int_0^\pi \int_{\gamma_1}^{\gamma_2} \mathrm{sin}^2\theta\,\gamma^{-\alpha}\,x\int_x^\infty K_{5/3}(\xi)\,d\xi\,d\gamma d\Omega_\theta.\\
\end{split}
    \label{eq:dW1}
\end{equation}
Upon comparison with the Eq. (\ref{eq:electrondist1}), $k = n_\mathrm{e,cr}\,A(\alpha,\gamma_1,\gamma_2)$. Further, assuming a radial profile for the magnetic field strength, Eq. (\ref{eq:dW1}) is re-written as
\begin{equation}
    \frac{dW(r)}{d\nu\,dt} = \frac{\sqrt{3}\,n_\mathrm{e,cr}\,A(\alpha,\gamma_1,\gamma_2)\,e^3B(r)}{8\pi m_\mathrm{e}c^2}\int_0^\pi \int_{\gamma_1}^{\gamma_2} \mathrm{sin}^2\theta\,\gamma^{-\alpha}\,x(r)\int_{x(r)}^\infty K_{5/3}(\xi)\,d\xi\,d\gamma d\Omega_\theta.
    \label{eq:dW2}
\end{equation}

We use a cluster sample (Table \ref{tab:catalogue}) where the synchrotron fluxes are scaled to a fixed observing frequency of 1.4 GHz. To get to the rest-frame emissivity following the Eq. (\ref{eq:dW2}) above, we use the cluster redshifts to convert to emission-frame frequencies. This is then integrated out to a fixed radius of $R_{500}$ to match the reported luminosity values. 


\subsection{Radial profiles of electrons and the magnetic field}

Finally, we describe the radial profiles used for matched-filtering the cluster SZ signal and model the synchrotron emissivity profiles. We assume the same pressure profiles for thermal and non-thermal electrons. Under the additional assumption of isothermal electrons (pseudo-temperature in the non-thermal case, Section \ref{sec:pseudoT}), the pressure profile also gives the density profile. The magnetic field strength is then related to this electron density profile by assuming that their energy densities will have the same radial dependence (see  \cite{Ensslin1998}).

\subsubsection{Pressure profile}
\label{sec:gnfw}
The spatial profile of the SZ effect is determined by the radial profiles of the Compton-y parameters, $y_\mathrm{th}$ (r) and $y_\mathrm{nth}$(r) [see Eq. (\ref{eq:di3})]. In order to model the radial profiles of $y_\mathrm{th}$(r) and $y_\mathrm{nth}$(r), we use the generalised Navarro-Frenk-White (GNFW) profile of the thermal electrons \cite{Nagai2007, Arnaud2010} with a fixed choice of the shape parameters. The only determining factors for the cluster pressure profile are then its mass and redshift.

The GNFW profile is used for modelling the distribution of thermal pressure within the ICM and is expressed as
\begin{equation}
    \frac{P(r)}{P_{500}} = \frac{P_0}{(c_{500}\frac{r}{R_{500}})^\gamma[1+(c_{500}\frac{r}{R_{500}})^\alpha]^{(\beta-\gamma)/\alpha}},
    \label{eq:pr}
\end{equation}
where 
\begin{equation}
R_{500} = \Big(\frac{3\,M_{500}}{4\pi\:500\rho_{\mathrm{crit}}}\Big)^{1/3}.
\end{equation}
Here, $\rho_{\mathrm{crit}}$ is the critical density at cluster redshift $z$, $c_{500}$ is the gas-concentration parameter, $P_0$ is the amplitude of pressure, and $\gamma$, $\alpha$, and $\beta$ describe the inner, intermediate and outer slopes of the profile. The slope parameter $\alpha$ should not be confused with the power-law index of the electron energy distribution [Eq. (\ref{eq:electrondist1})]. The parameters ($c_{500}$, $\gamma$, $\alpha$, $\beta$) are referred to as shape parameters. We adopt 
\begin{equation}
    P_{500} = 1.65\times10^{-3}\:h(z)^{8/3}\times \Bigg[\frac{M_{500}}{3\times10^{14}h_{70}^{-1} \: M_\odot}\Bigg]^{2/3 +\alpha_\mathrm{p}+\alpha_\mathrm{p}'(r)}h_{70}^2 \: \mathrm{keV}\: \mathrm{cm}^{-3}
    \label{eq:p500},
\end{equation}
presented in \cite{Arnaud2010} with their best-fit parameters of $P_0=8.403\,h_{70}^{-3/2}$, $\alpha_\mathrm{p}=0.120$, $\alpha=1.051$, $\beta=5.4905$, $\gamma=0.3081$, $c_{500} = 1.177$ and $\alpha_\mathrm{p}'(r)=0$. In Eq. (\ref{eq:p500}), $h(z)=\frac{H(z)}{H_0}$ is the reduced Hubble parameter at cluster redshift $z$, and $h_{70} = \frac{H_0}{70\,\mathrm{km}\,\mathrm{s}^{-1}\mathrm{Mpc}^{-1}}$.
\paragraph{}
The GNFW pressure profile is then integrated along the line-of-sight (los) to compute the radial profile of the Compton-y parameter, 
\begin{equation}
    y(r) = \frac{\sigma_T}{m_\mathrm{e}c^2}\int_{\mathrm{los}}P_\mathrm{e}(r)\: dl,
    \label{eq:yr}
\end{equation}
where $P_\mathrm{e}(r)$ is described by Eqs. (\ref{eq:pr}) and (\ref{eq:p500}). This projection is done numerically with the assumption of spherical symmetry, and the resulting $y$-profile (for each individual cluster) is taken as the template for optimally extracting the cluster tSZ$+$ntSZ signal via matched filtering.

\subsubsection{Magnetic field}
In order to compute the radial profile of the synchrotron emission (Eq. (\ref{eq:dW2})) for a given radio halo, we need radial profiles of the non-thermal electrons and the magnetic field in the ICM. With the relation $P_\mathrm{e}(r) = n_\mathrm{e}(r)\langle k_\mathrm{B}T_\mathrm{e}\rangle$, we assume the deprojected GNFW profile for $n_\mathrm{e}(r)$, and the corresponding distribution of the magnetic field is
\begin{equation}
    B(r) = B_0\Bigg(\frac{n_\mathrm{e}(r)}{n_{\mathrm{e,0}}}\Bigg)^{0.5},
    \label{eq:Br}
\end{equation}
where $B_0$ and $n_{\mathrm{e,0}}$ are the central magnetic field strength and electron number density, respectively. As discussed in Section \ref{sec:theory}, this radial dependence follows from an energy equipartition argument wherein the magnetic field energy density and the relativistic electron density have the same radial scaling \cite{Ensslin1998}. Observational evidence of this power-law dependence has been demonstrated by \cite{Murgia2004, Bonafede2010}. While the exact value of the power-law index is not critical for our analysis, a profile where the magnetic field strength scales down with radius is necessary to compute a realistic estimate of the synchrotron power.

\begin{figure}[ht]
    \centering
    \includegraphics[width=0.8\textwidth]{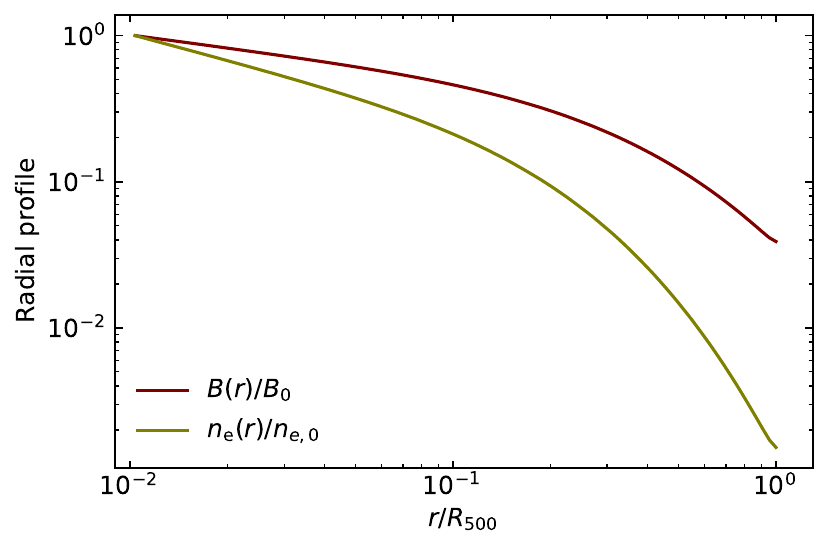}
    \caption{Radial profiles of the electron number density (which traces the GNFW pressure profile) and the magnetic field strength as a function of $r/R_{500}$. Both profiles are normalized to unity to highlight the diverging rate of radial fall-off.}
    \label{fig:neBprofile}
\end{figure}

The radial profiles of the $n_\mathrm{e}(r)$ and $B(r)$ are plotted in Figure \ref{fig:neBprofile}. We find that the magnetic field profile is significantly flatter than the electron density profile (the latter having identical shape as the thermal pressure profile under the assumption of isothermality). This translates into different factors of improvement on the electron number density and magnetic field constraints, when a future experiment with improved sensitivities is considered. We fit for the central magnetic field strength via Eqs.\ \eqref{eq:Br} and \eqref{eq:dW2}, and present the estimated central and volume-averaged magnetic field strength for each of the non-thermal electron models considered in Section \ref{sec:results}.

\section{Data and simulations}
\label{sec:data}
Observations in the mm/sub-mm wavelength range are necessary to exploit the difference in spectral shapes of the tSZ and ntSZ signals. The zero-crossing frequency, which is dependent on the non-thermal electron momenta distribution, also lies in this range. CMB surveys offer data in exactly this regime. 
\subsection{Radio halo cluster sample}
Our sample of GCs hosting radio halos are compiled from \cite{vanweeren2019}. 62 such radio halos are selected and their coordinates, $M_{500}$ and redshift estimates are obtained from the second \textit{Planck} catalogue of SZ sources \cite{Planck2015}. The synchrotron radiation flux measurements at 1.4 GHz for a sub-sample of 32 GCs \cite{Yuan2015, Gennaro2021} and the associated spectral index of the power-law describing the synchrotron emission are obtained from literature. These characteristics of our sample of GCs are tabulated in Table \ref{tab:catalogue}. A mean synchrotron power of $1.54\times10^{31}\,\mathrm{erg\,s}^{-1}\mathrm{Hz}^{-1}$ is assumed to obtain constraints on the magnetic field strength in GCs.

\begin{longtable}{cccc}
\hline \multicolumn{1}{c}{\textbf{Cluster}} & \multicolumn{1}{c}{\textbf{$z$}} & \multicolumn{1}{c}{\textbf{$M_{500}$}} &
\multicolumn{1}{c}{\textbf{log$_{10}$($P_{1.4\,\mathrm{GHz}}$})} \\ 
& & ($\times 10^{14} \mathrm{M}_\odot$) & \\
\hline 
\endfirsthead
\multicolumn{4}{c}
{{\bfseries \tablename\ \thetable{} -- continued from previous page}} \\
\hline \multicolumn{1}{c}{\textbf{Cluster}} & \multicolumn{1}{c}{\textbf{$z$}} & \multicolumn{1}{c}{\textbf{$M_{500}$}} &
\multicolumn{1}{c}{\textbf{log$_{10}$($P_{1.4\,\mathrm{GHz}}$})} \\ 
& & ($\times 10^{14} \mathrm{M}_\odot$) & \\
\hline 
\endhead
\hline \multicolumn{4}{r}{{Continued on next page}} \\ \hline
\endfoot
\endlastfoot
        Coma&0.023&7.165297&-0.19$\pm$0.04\\
        A3562&0.049&2.443&-0.95$\pm$0.05\\
        A754&0.054&6.853962&-0.24$\pm$0.03\\
        A2319&0.056&8.735104&0.24$\pm$0.02\\
        A2256&0.058&6.210739&-0.08$\pm$0.01\\
        A399&0.072&5.239323&-0.7$\pm$0.06\\
        A401&0.074&6.745817& \\
        A2255&0.081&5.382814&-0.06$\pm$0.02\\
        A2142&0.089&8.771307&-(0.72$\pm$1.22)\\
        A2811&0.108&3.647853&  \\
        A2069&0.115&5.30745&  \\
        A1132&0.137&5.865067&-(0.79$\pm$1.09)\\
        A3888&0.151&7.194754&0.28$\pm$0.69\\
        A545&0.154&5.394049&0.15$\pm$0.02\\
        A3411-3412&0.162&6.592571&-(0.57$\pm$1.0)\\
        A2218&0.171&6.585151&-0.41$\pm$0.01\\
        A2254&0.178&5.587061& \\
        A665&0.182&8.859059&0.58$\pm$0.02\\
        A1689&0.183&8.768981&-0.06$\pm$0.15\\
        A1451&0.199&7.162284&-(0.19$\pm$1.15)\\
        A2163&0.203&16.116468&1.24$\pm$0.01\\
        A520&0.203&7.80038&0.26$\pm$0.02\\
        A209&0.206&8.464249&0.24$\pm$0.02\\
        A773&0.217&6.847479&0.22$\pm$0.05\\
        RXCJ1514.9-1523&0.223&8.860777&0.14$\pm$0.1\\
        A2261&0.224&7.77852&-(0.17$\pm$1.15)\\
        A2219&0.228&11.691892&1.06$\pm$0.02\\
        A141&0.23&5.672555&  \\
        A746&0.232&5.335297&0.43$\pm$0.11\\
        RXCJ1314.4-2515&0.247&6.716546&-(0.17$\pm$0.62)\\
        A521&0.248&7.255627&0.07$\pm$0.04\\
        A1550&0.254&5.877626&  \\
        PSZ1G171.96-40.64&0.27&10.710258&0.58$\pm$0.05\\
        A1758&0.28&8.217337&0.72$\pm$0.11\\
        A697&0.282&10.998416&0.08$\pm$0.04\\
        RXCJ1501.3+4220&0.292&5.869359& \\ 
        Bullet&0.296&13.100348&1.16$\pm$0.02\\
        A2744&0.308&9.835684&1.21$\pm$0.02\\
        A1300&0.308&8.971329&0.58$\pm$0.16\\
        RXCJ2003.5-2323&0.317&8.991968&1.03$\pm$0.03\\
        A1995&0.318&4.924279&0.1$\pm$0.08\\
        A1351&0.322&6.867679&1.01$\pm$0.06\\
        PSZ1G094.00+27.41&0.332&6.776592&0.58$\pm$0.02\\
        PSZ1G108.18–11.53&0.335&7.738726&  \\
        MACSJ0949.8+1708&0.383&8.23875& \\
        MACSJ0553.4–3342&0.407&8.772141 & \\
        MACSJ0417.5–1154&0.443&12.250381 & \\
        MACSJ2243.3–0935&0.447&9.992374 & \\
        MACSJ1149.5+2223&0.544&10.417826 & \\
        MACSJ0717.5+3745&0.546&11.487184& \\
        ACT-CLJ0102–4915&0.87&10.75359 & \\
        AS1121&0.358&7.193831&  \\
        ZwCl0634+4750803&0.174&6.652367&-(0.51$\pm$1.69)\\
        PLCKG004.5-19.5&0.54&10.356931&  \\
        CL0016+16&0.5456&9.793704&0.76$\pm$0.07\\
        PLCKESZG285–23.70&0.39&8.392523 & \\
        RXCJ0256.5+0006&0.36&5.0 & \\
        MACSJ1752.0+4440&0.366&4.3298&1.1$\pm$0.03\\
        A800&0.2472&3.1464&  \\
        CL1446+26&0.37&2.70015& \\ 
        CIZAJ2242.8+5301&0.192&4.0116&1.16$\pm$0.05\\
        MACSJ0416.1–2403&0.396&4.105336&  \\
\hline
\caption{Cluster identifiers, redshift ($z$), mass ($M_{500}$) and synchrotron power at 1.4 GHz (in $10^{24}\,$W/Hz) that are used in this work.}
\label{tab:catalogue}
\end{longtable}

\subsection{{\it Planck} all-sky maps}
The \textit{Planck} satellite observed the sky for four years with two instruments. The Low-frequency Instrument (LFI) was sensitive in the 30 -- 70 GHz range \cite{Planck2018LFI} and the High-frequency Instrument (HFI) was sensitive in the 100 -- 857 GHz range \cite{Planck2018HFI}. We used the 2018 release of the {\it Planck} all-sky multi-frequency maps \cite{Planck2018overview}. 70 GHz maps from the LFI and maps from all six bands from the HFI are used. The maps are available in \texttt{HEALPix}\footnote{\url{http://healpix.sourceforge.net}} format \cite{Gorski2005} with N\textsubscript{side} = 2048 for the HFI channels and N\textsubscript{side} = 1024 for the LFI channels. We chose to work in units of surface brightness (MJy$\,\mathrm{sr}^{-1}$) and this required the 70 -- 353 GHz maps, which are originally available in units of K\textsubscript{CMB}, to be converted to surface brightness maps using the Unit Conversion - Colour Correction (UC-CC)\footnote{\url{https://wiki.cosmos.esa.int/planckpla2015/index.php/UC\_CC\_Tables}} tables. The resolution of the maps and the respective UC values are tabulated in Table \ref{tab:ucres}.

\begin{table}[ht]
    \centering
    \begin{tabular}{ccc}
    \hline
         Frequency band & FWHM & UC\\
         (GHz) & (arcmin) & ($\mathrm{MJy\,sr}^{-1}\,\mathrm{K_{CMB}}^{-1}$)\\
         \hline
         70 & 13.31& 129.187\\
         100 & 9.68 & 244.096\\
         143 & 7.30 & 371.733\\
         217 & 5.02 & 483.687\\
         353 & 4.94 & 287.452\\
         545 & 4.83 & 58.036\\
         857 & 4.64 & 2.2681 \\
         \hline
    \end{tabular}
    \caption{Resolution of \textit{Planck}'s 70 GHz LFI band and all HFI frequency bands. The unit conversion coefficients are used to convert the 70 -- 353 GHz maps from units of $\mathrm{K_{CMB}}$ to $\mathrm{MJy\,sr}^{-1}$.}
    \label{tab:ucres}
\end{table}

\subsection{Simulated microwave sky maps}
\label{subsec:simsky}
In order to quantify the constraining power of upcoming CMB experiments in obtaining upper limits on the non-thermal electron number density, an estimate of the noise covariance matrix is required. We first simulate the microwave sky maps at the observing frequencies and convolve them with a Gaussian beam (Table \ref{tab:sensitivities}).
\paragraph{}
The simulated maps comprise of the following components:
\begin{itemize}
    \item Galactic foregrounds (dust, synchrotron, anomalous microwave emission, and free-free emission) which are simulated using the Python Sky Model (PySM) software \cite{PySM}.
    \item Cosmic Infrared Background (CIB), CMB, radio point sources, tSZ, and kinetic SZ (kSZ; \cite{Sunyaev1980}) components which are simulated using the Websky extragalactic CMB simulations \cite{Websky}.
\end{itemize}
Finally, white noise with variance given by the sensitivites  of the instruments \cite{Choi2019, SimonsObservatory2018} (listed in Table \ref{tab:sensitivities}) are added to the maps to represent the detector noise and any residual atmospheric noise. Since we consider small areas of the sky, the dominant noise component is the white noise and we choose to ignore the $1/f$ noise component. \texttt{ccatp\_sky\_model}\footnote{\url{https://github.com/MaudeCharmetant/CCATp_sky_model}} \texttt{Python} package, which incorporates the PySM and Websky simulations, is used to simulate the microwave sky maps in this work.
\begin{table}[ht]
        \centering
    \begin{tabular}{ccc}
    \hline
    Frequency band & FWHM & Sensitivity\\
     (GHz) & (arcmin) & ($\mu$K-arcmin)\\
    \hline
    27 & 7.4 & 71 \\
    39 & 5.1 & 36 \\
    93 & 2.2 & 8 \\
    145 & 1.4 & 10 \\
    225 & 1.1 & $\left[ 22^{-2} + 15^{-2} \right]^{-1/2}$ \\
    280 & 1.1 & $\left[ 54^{-2} + 27^{-2} \right]^{-1/2}$ \\
    350 & 1.1 & 105 \\
    405 & 1.1 & 372 \\
    860 & 1.1 & $5.75\times 10^5$ \\
    \hline
        \end{tabular}
            \caption{Sensitivities of FYST, SO and the combined sensitivities of SO+FYST configuration. For the common frequencies between these two experiments (225 and 280 GHz) the noise values are added in quadrature (following inver-variance weighting), where the first number inside the parenthesis corresponding to the SO, and the second value corresponding to FYST.
            }
    \label{tab:sensitivities}
\end{table}
\section{Methods}
\label{sec:fit}

We first extract $10^\circ \times 10^\circ$ cluster fields from the all-sky multi-frequency maps centred around the coordinates of each of the GC in our sample using the \texttt{gnomview()} function of the \texttt{healpy}\footnote{\url{https://healpy.readthedocs.io/en/latest/index.html}} \cite{healpy} module. To improve the Signal-to-Noise Ratio (SNR) of the SZ effect and minimize the amplitude of contaminants in the cluster fields, we employ the following methods:
\begin{enumerate}
    \item Matched-filtering (MF): This method is used to minimize the noise and other astrophysical contaminants to optimally extract the cluster signal, assuming a fixed spatial template.
    
    \item Stacking: Stacking cluster fields ensures amplification of the tSZ and ntSZ signals by averaging the uncorrelated noise, and minimizes the kSZ signal from individual clusters.
\end{enumerate}
In the following subsections we discuss these methods in detail.

\subsection{Matched-filtering method}
Matched filters are designed to optimally extract a signal in the presence of Gaussian noise and have been shown to be effective for extracting cluster SZ signals (e.g. \cite{Haehnelt1996, Melin2006, Erler2018, Zubeldia2021}). Presence of small non-Gaussian noise components will not cause a bias but the solution may not be optimal \cite{Melin2006}. 
In practice, setting up a matched filter is extremely easy as it only requires that we know the spatial template of the emitting sources. In the flat-sky approximation, this filter function (a vectorized map) can be written as the following in the Fourier space \cite{Schaefer2006, Erler2019}:
\begin{equation}
\boldmath{\Psi} = \left[ \mathbf{\tau}^\mathrm{T} \mathbf{C}^{-1} \mathbf{\tau}\right]^{-1} \mathbf{\tau}\mathbf{C}^{-1},
\end{equation}
where $\boldsymbol{\tau}$ is the Fourier transform of the 2D $y$-profile model, and $\mathbf{C}$ is the azimuthally-averaged noise power spectrum of the unfiltered map. We use the publicly available \texttt{PYTHON} implemention of MF, called \texttt{PyMF}\footnote{\url{https://github.com/j-erler/pymf}} \cite{Erler2019}, to filter the $10^\circ \times 10^\circ$ maps. The noise power spectra are computed from the same fields after masking the GCs, or, in the case of forecasts, from random empty fields.

\subsection{Stacking}
\begin{figure}[ht]
    \centering
    \includegraphics[width=1.0\textwidth]{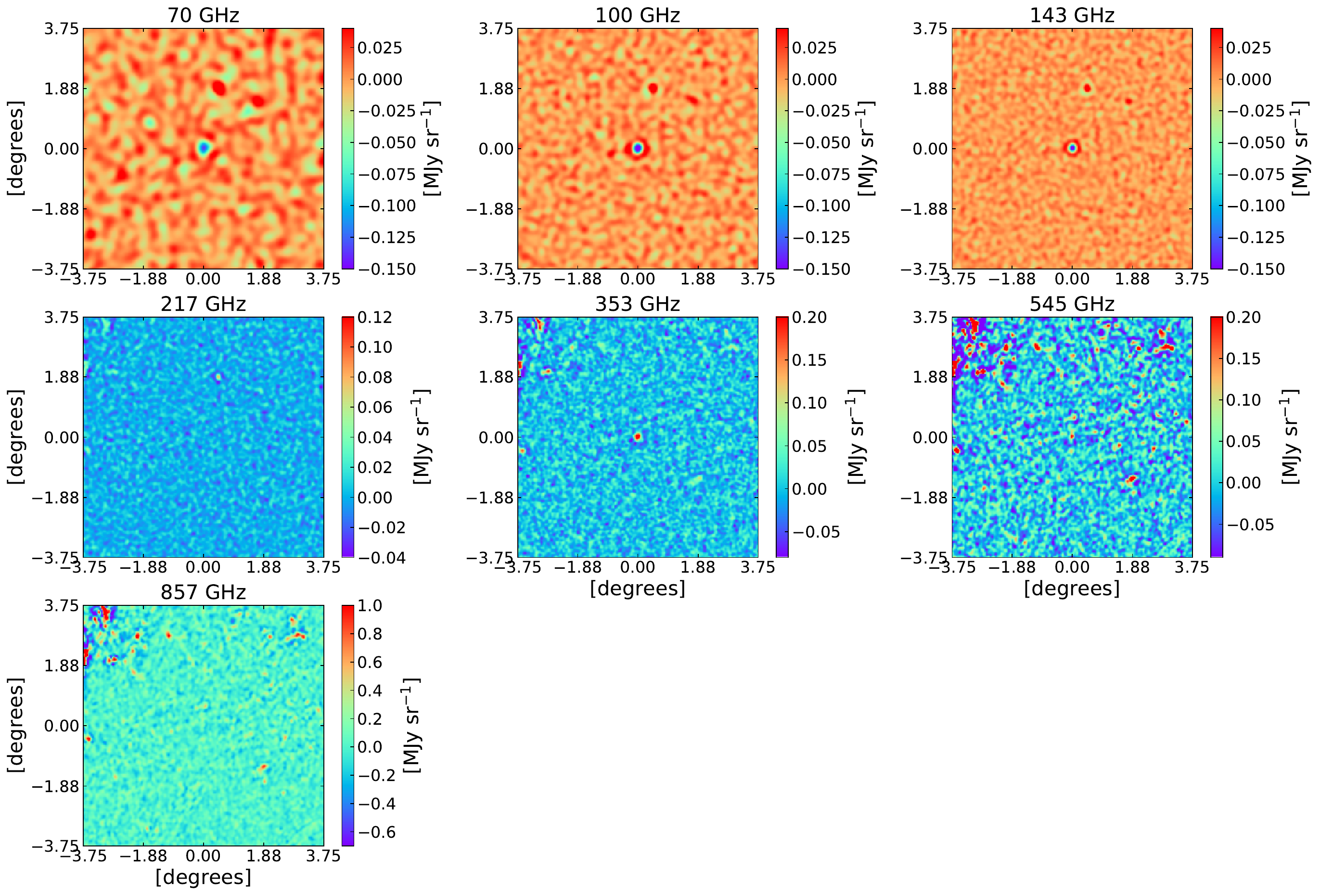}
    \caption{$7.5^\circ \times 7.5^\circ$ stacked matched-filtered maps of the 62 galaxy clusters in the sample for each of the 70 GHz \textit{Planck} LFI and HFI channels. The SZ effect signal is clearly seen in the centre. There is still some residual dust emission visible in the HFI maps. The color scale is intentionally set differently for each map in order to enhance any features in the map.}
    \label{fig:stackedmf}
\end{figure}

Matched-filtered cluster fields are stacked to obtain an \textit{average} filtered map at each observed frequency. Since the noise properties are practically Gaussian after filtering, this leads to a suppression of noise by roughly a factor of $1/\sqrt{62}$, where 62 is the number of clusters in our radio halo sample. Stacking also has the additional advantage of suppressing the kSZ signal by the same factor, which acts as a random source of noise at the cluster location. 
The stacked matched-filtered maps are shown in Figure \ref{fig:stackedmf}.

\begin{figure}[ht]
    \centering
    \includegraphics[width=0.8\textwidth]{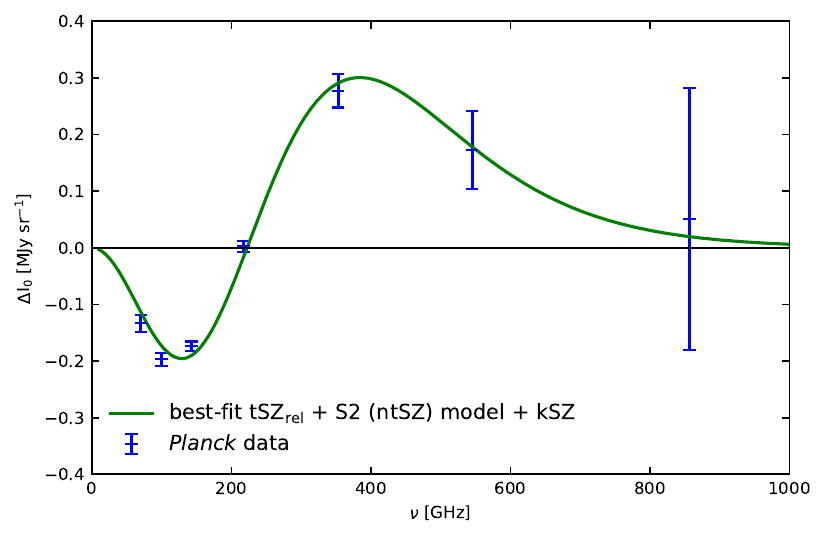}
    \caption{Spectrum of the SZ effect extracted from the stacked matched-filtered maps of the \plk data. The values correspond to the central pixel value in the stacked matched-filtered maps. The error bars correspond to the extent of the variance of foregrounds in the maps.}
    \label{fig:extractedspec}
\end{figure}

The amplitude of the SZ signal at each frequency channel is in the central pixel of the stacked matched-filtered map, and this value is extracted to obtain a spectrum of the SZ effect. The extracted spectrum is displayed in Figure \ref{fig:extractedspec}. It shows the characteristic shape of the SZ effect with a decrement in specific intensity at frequencies $\nu <$ 217 GHz and an increment at higher frequencies. The error bars correspond to the variance of astrophysical emission in the stacked matched-filtered maps. The uncertainties for the high-frequency channels are larger as the mean contribution from dust emission is still prominent in the maps and the HFI at $\nu = 353,\, 545\,$ and 857 GHz, in general, are relatively noisy.

\subsection{Spectral fitting}
The extracted amplitude of the SZ signal can be decomposed into the distortions due to $\mathrm{tSZ_{rel}}$, kSZ and ntSZ effects as\footnote{The subscript 0 refers to the fact that the amplitudes correspond to the central pixel in the maps.}
\begin{equation}
\begin{split}
    \Delta I_{\mathrm{0,\nu}} &= \delta i_{\mathrm{0,\nu}}^{\mathrm{th}} + \delta i_{\mathrm{0,\nu}}^{\mathrm{nth}} + \delta i_{\mathrm{0,\nu}}^{\mathrm{kSZ}}\\
    &=\Bigg[\Bigg(\int_{p_1}^{p_2}\int_{-s_{\mathrm{m}}(p_1)}^{s_{\mathrm{m}}(p_2)} f_\mathrm{e,th}(p;\Theta) K(e^\mathrm{s};p)\,e^\mathrm{s}\: \frac{(x/e^\mathrm{s})^3}{(e^{x/e^\mathrm{s}}-1)}\: ds\: dp\Bigg) - \frac{x^3}{e^x-1}\Bigg] I_0\frac{m_\mathrm{e} c^2}{k_\mathrm{B}T_\mathrm{e}}\,y_0^\mathrm{th} \\
    &+ \Bigg[\Bigg(\int_{p_1}^{p_2}\int_{-s_\mathrm{m}(p_1)}^{s_\mathrm{m}(p_2)}f_\mathrm{e}(p)\:K(e^\mathrm{s};p)\,e^\mathrm{s}\: \frac{(x/e^\mathrm{s})^3}{(e^{x/e^\mathrm{s}}-1)}\,ds\: dp \Bigg) \,-\, \frac{x^3}{e^x-1}\Bigg]\, I_0\frac{m_\mathrm{e}c^2}{\langle k_\mathrm{B}\tilde{T}_\mathrm{e}\rangle}\,y_0^\mathrm{{nth}} \\
    & -I_0\: \frac{x^4e^x}{(e^x-1)^2}\, y_0^\mathrm{kSZ},  
    \label{eq:excomp1}
\end{split}
\end{equation}

where $\Delta I_\mathrm{0,\nu}$ is the amplitude of the SZ effect signal from stacked matched-filtered map of frequency $\nu$, $\delta i_{\mathrm{0,\nu}}^\mathrm{th}$, $\delta i_{\mathrm{0,\nu}}^\mathrm{kSZ}$ and $\delta i_{\mathrm{0,\nu}}^\mathrm{nth}$ are distortions due to the tSZ$_\mathrm{rel}$, kSZ and ntSZ effects, respectively; $f_\mathrm{e,th}(p;\Theta)$ is the Maxwell-J\"{u}ttner distribution used to describe the thermal distribution of electrons in terms of the normalized thermal energy parameter, $\Theta = \frac{k_\mathrm{B}T_\mathrm{e}}{m_\mathrm{e}c^2}$, and $y_0^\mathrm{kSZ}$ is analogous to $y_0^\mathrm{th}$.  Appendix \ref{sec:modelsz} can be referred to for more information on how we compute the $\mathrm{tSZ_{rel}}$ and kSZ spectra. We thus fit the extracted SZ spectrum with a three-component model consisting of the tSZ$_\mathrm{rel}$, kSZ and ntSZ signals, using the MCMC sampling method.  While fitting this three-component model to \textit{Planck} data, we use bandpass corrected spectra (a description of which can be found in Appendix \ref{sec:bandpass}). 

The shape of the $\mathrm{tSZ_{rel}}$ spectrum is fixed by using a single $T_\mathrm{e}$ to represent our stack of clusters and fit only for the amplitude, $y_\mathrm{0,th}$. The spectral distortion due to $\mathrm{tSZ_{rel}}$ is described in Appendix \ref{app:rSZ}. Since we work with the stacked signal for spectrum fitting, we adopt a single, median temperature from all the clusters for $T_\mathrm{e}$, where individual cluster temperatures are obtained from a mass-temperature relation as given in \cite{Reichert2011}. The median temperature (energy) is approximately 8 keV and is used for computing the relativistic corrections to the tSZ signal. 

The $y_\mathrm{{kSZ}}$ is marginalized over by drawing $v_\mathrm{{pec}}$ from a Gaussian distribution of zero mean and a standard deviation corresponding to the expected line-of-sight velocity after stacking. This marginalization amplitude is $100/\sqrt{62}$ \,kms$^{-1}$ in each step of the chain. We also fix the shape of the ntSZ spectrum by assuming specific values for $p_1$ and $p_2$ and fit for the amplitude of the ntSZ effect, $y_{\mathrm{0,nth}}$. Finally, the posterior probability distributions of $y_{\mathrm{th}}$ and $y_\mathrm{{nth}}$ are estimated for four different models of the non-thermal electron momentum distribution. To fit the stacked spectrum we need a frequency-to-frequency noise covariance. In the case of \textit{Planck} data, the covariance matrix is computed empirically from the stacked matched-filtered maps as
\begin{equation}
    C_{ij} \equiv \frac{1}{N_\mathrm{pix}-1}\sum_{p=1}^{N_\mathrm{pix}} (I_{i}(p)-\overline{I}_{i})(I_{j}(p)-\overline{I}_{j}),
    \label{eq:covmatrix}
\end{equation}
where $N_\mathrm{pix} = 1600$ denotes the number of pixels of pixel size $1.5'$ in a $10^\circ \times10^\circ$ field and $\mathbf{I}(p)$ denotes the value of pixel $p$ in intensity map \textbf{I}. The covariance matrix is estimated by masking the cluster region in the stacked matched-filtered maps. In the case of SO$+$FYST forecasts, the frequency covariance is computed in a similar way from randomly located empty fields.

\begin{figure}
    \centering
    \includegraphics[width=1.0\textwidth]{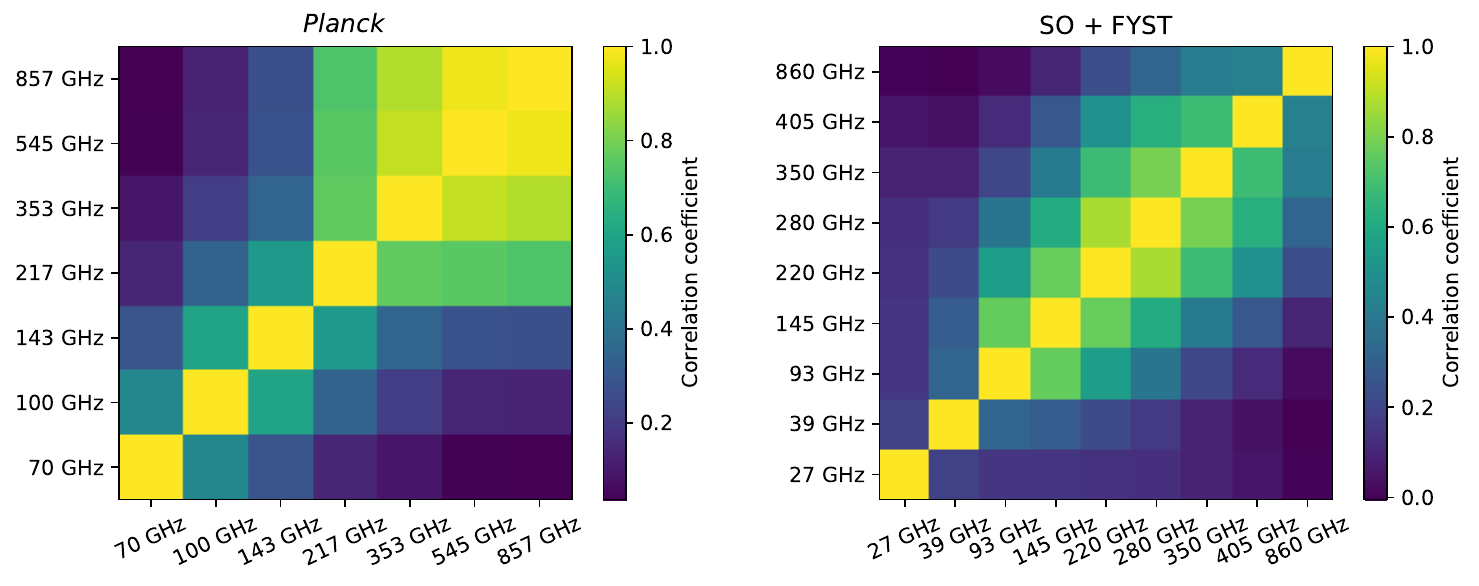}
    \caption{The spectral correlation between the different frequency maps from \textit{Planck} data (\textit{left}) and simulated frequency maps with SO+FYST sensitivities (\textit{right}).}
    \label{fig:correlationcoeff}
\end{figure}

\paragraph{}
For our forecasts, we follow a similar procedure of computing the noise covariance matrices using stacked matched-filtered maps extracted from simulated maps described in Section \ref{subsec:simsky}. Specifically, 100 $10^\circ \times10^\circ$ fields centered around coordinates sampled from a uniform distribution are extracted from a full-sky simulated map, and a mean of these 100 fields is computed to represent foregrounds in one simulated cluster field. We then compute 62 such cluster fields, perform matched-filtering and stack the matched-filtered fields to get one simulated stacked matched-filtered cluster field. This procedure is performed at each observing frequency. These simulated stacked fields are then used to compute the noise covariance matrix described in Eq. (\ref{eq:covmatrix}). Figure \ref{fig:correlationcoeff} shows the correlation matrices computed from \textit{Planck} data and the simulated maps with the SO+FYST configuration.

\section{Results}
\label{sec:results}

Due to the presence of the dominant tSZ$_\mathrm{rel}$ effect and the constraining power of the sensitivities of \textit{Planck}, we are able to obtain upper limits on the $y_\mathrm{nth}$ and $n_\mathrm{e,nth}$, and further, lower limits on the magnetic field strength.  

\subsection{Current constraints from the {\it Planck} data}
The constraints on the amplitude of the tSZ$_\mathrm{rel}$ and ntSZ effects are shown in Table \ref{tab:results1}. With the assumption of isothermality and a median $k_\mathrm{B}T_\mathrm{e} = 8.0\,$keV for the stack of GCs, the $y_0^{\mathrm{th}}$ is well constrained by \textit{Planck} data. However, for $y_0^\mathrm{nth}$, we are only able to obtain upper limits with the data. Models of electron momenta that assume higher energies result in higher upper limits for $y_\mathrm{nth}$. The average electron number density remains consistent for all models. A large variation in constraints (for a fixed synchrotron flux density at 1.4 GHz) on the central and volume-averaged magnetic field strengths is observed.
\begin{table}[ht]
        \centering
    \begin{tabular}{ccccccc}
    \hline
    Model & Obs & $y_0^\mathrm{th}$& $y_0^\mathrm{nth}$& $\Bar{n}_\mathrm{e,nth}$ & $\Bar{B}$ & B$_0$\\
    &  & ($\times 10^{-4}$) & ($\times 10^{-4}$) & ($\times 10^{-6}$ cm$^{-3}$)&($\mu$G)&($\mu$G)\\
    \hline
         & & & & & &\\
         S1& \textit{Planck}& $1.83^{+0.09}_{-0.10}$ & $<4.81$ & $<2.06$& $>0.24$ & $>1.05$\\
         & & & & & &\\
        S2& \textit{Planck} & $1.82^{+0.09}_{-0.10}$ & $<48.61$ & $<2.08$ & $>0.02$ & $>0.08$\\
        \hline
        & & & & & &\\
         B1& \textit{Planck} & $1.82^{+0.09}_{-0.10}$ & $<23.83$ & $<2.09$& $>0.03$ & $>0.15$\\
        & & & & & & \\
        B2& \textit{Planck} & $1.82^{+0.09}_{-0.10}$ & $<77.70$ & $<2.05$& $>0.01$ & $>0.03$\\
        & & & & & &\\
        \hline
        \end{tabular}
            \caption{$y_0^\mathrm{th}$ and upper limits on $y_0^\mathrm{nth}$ obtained from the \textit{Planck} data. The number density and magnetic field strength are volume-averaged quantities within the $r_{500}$. For a central magnetic field strength of 1$\mu$G, and $\alpha=3.6$ as the index of the power-laws describing the electron momentum distribution, the volume-averaged value is $0.23\,\mu$G.}
    \label{tab:results1}
\end{table}
\subsection{Upcoming constraints from SO and FYST}
\begin{figure}[ht]
    \centering
    \includegraphics[width=0.8\textwidth]{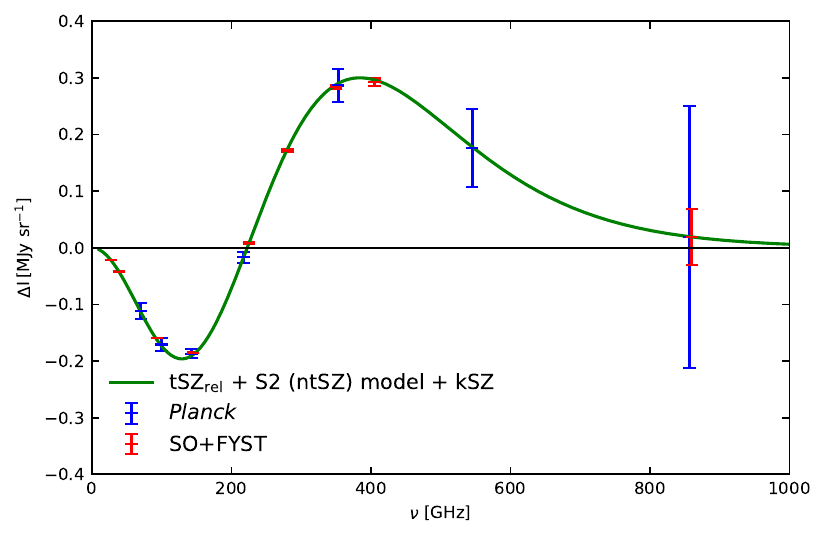}
    \caption{A comparison of noise variance expected for SO+FYST for a stack of 62 fields with \textit{Planck}. Observations of more radio halos could further improve the constraints estimated in this study.}
    \label{fig:noise_forecast}
\end{figure}

Assuming the sensitivities of SO+FYST configuration tabulated in Table \ref{tab:sensitivities}, we check for the constraining capabilities of upcoming CMB experiments on the amplitude of ntSZ effect. Further, the lower limits on magnetic field strength are computed and tabulated in Table \ref{tab:results2}. The expected variance in the measurement of the SZ spectrum for a stack of 62 cluster fields is plotted in Figure \ref{fig:noise_forecast} compared to the variance from \textit{ Planck} data. The error bars are significantly smaller at all frequencies due to the combined sensitivities of SO and FYST.
\begin{table}[ht]
        \centering
    \begin{tabular}{cccccc}
    \hline
    Model & Obs & $y_0^\mathrm{nth}$& $\Bar{n}_\mathrm{e,nth}$ & $\Bar{B}$ & B$_0$\\
    &  & ($\times 10^{-4}$) & ($\times 10^{-6}$ cm$^{-3}$)&($\mu$G)&($\mu$G)\\
    \hline
         & & & &\\
         S1& SO+FYST& $<1.10$ & $<0.47$ & $>0.46$ & $>2.0$\\
         & & & & \\
        S2& SO+FYST & $<11.37$ & $<0.49$ & $>0.04$ & $>0.16$\\
        \hline
        & & & &\\
         B1& SO+FYST & $<5.07$ & $<0.47$ & $>0.07$ & $>0.28$\\
        & & & &\\
        B2& SO+FYST & $<17.29$ & $<0.48$ & $>0.02$ & $>0.07$\\
        & & & &\\
        \hline
        \end{tabular}
            \caption{Same as Table \ref{tab:results1} but for SO+FYST.}
    \label{tab:results2}
\end{table}

\begin{figure}[ht]
    \centering
    \includegraphics[width=1.0\textwidth]{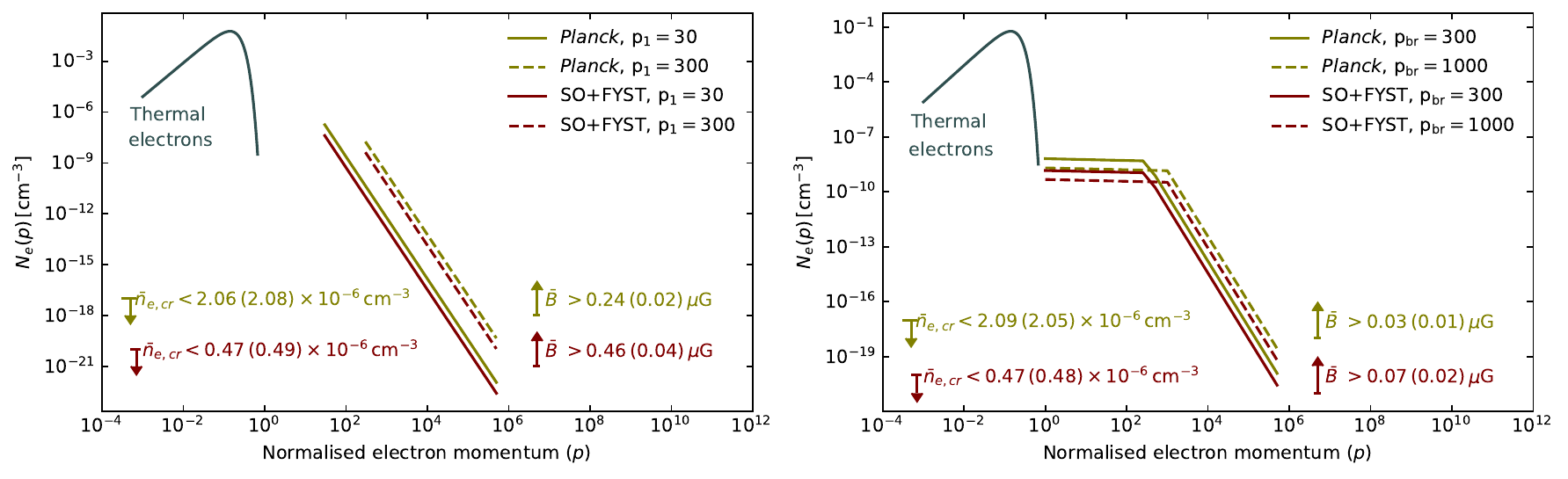}
    \caption{Thermal and non-thermal populations of scattering electrons in the ICM. $k_\mathrm{B}T_\mathrm{e} = 8.0$ keV is assumed for the distribution of thermal electrons. Left: Power-law distributions of electrons are assumed for the non-thermal electrons with constraints estimated from \textit{Planck} data and SO+FYST configuration. Right: Broken power-law is assumed for the non-thermal electrons with constraints from \textit{Planck} and SO+FYST configuration. Full and broken lines correspond to different minimum (broken) momenta for power-law (broken) distributions.}
    \label{fig:fullnedist}
\end{figure}
\begin{figure}[ht]
    \centering
    \includegraphics[width=1.\linewidth]{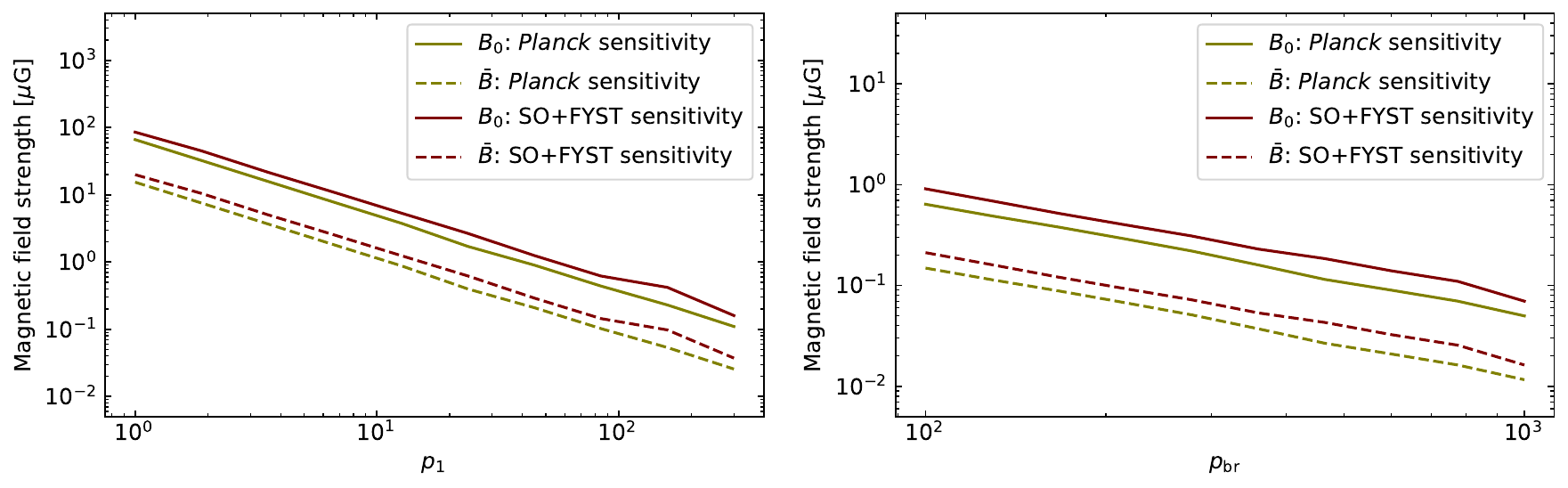}
    \caption{Lower limits on the central magnetic field strength ($B_0$) and volume-averaged field strength ($\bar{B}$) for different values of $p_1$ (\textit{left}) and $p_\mathrm{br}$ (\textit{right}) for \plk (in \textit{olive}) and SO+FYST (in \textit{maroon}) sensitivities.}
    \label{fig:p1-pbr-B}
\end{figure}

\paragraph{}
We also estimate the central and volume-averaged magnetic field strength assuming a mean synchrotron power with \plk and SO+FYST sensitivities as a function of $p_1$ for single power-law and $p_\mathrm{br}$ for the broken power-law models describing the non-thermal electron momentum distributions [Eqs. \eqref{eq:electrondist1} and \eqref{eq:electrondist2}]. The estimated lower limits on $B_0$ and $\bar{B}$ are plotted in Figure \ref{fig:p1-pbr-B}. The dependence of the synchrotron emission on the electron momentum is determined by the synchrotron kernel $x\,\int_x^\infty K_{5/3} (\xi)\,d\xi$ in Eqs. \eqref{eq:synch1}, \eqref{eq:dW1} and \eqref{eq:dW2}. Depending on the magnetic field strength (and thus the critical frequency, Eq.\ \eqref{eq:nu-c}), the synchrotron kernel probes different regions of the non-thermal electron momentum distribution. For synchrotron emission at 1.4 GHz, it is only the higher-momentum tail of the distribution that contributes to the emission. When we consider lower $p_1$ (or $p_\mathrm{br}$) and a fixed synchotron power, higher magnetic field strengths will be estimated as there are lower number of high-momentum non-thermal electrons emitting synchrotron radiation at 1.4 GHz and this is evident in Figure \ref{fig:p1-pbr-B} wherein $B_0$ increases with decreasing $p_1$ and $p_\mathrm{br}$.

\section{Discussion and conclusions}
\label{sec:summary}

The amplitude of the $\mathrm{tSZ_{rel}}$ effect is well constrained by \textit{Planck} under the assumption of an isothermal ICM, and simultaneously, for the first time, upper limits on the ntSZ amplitude and non-thermal electron number density for different models of the non-thermal electron momenta are obtained. The resulting number densities of the different populations of electrons in the ICM, as derived from the constraints, are plotted in Figure \ref{fig:fullnedist}. While $y_\mathrm{nth}$ and $N_\mathrm{e}(p)$ are sensitive to the choice of models of $f_\mathrm{e}(p)$, there is no significant variation in the volume-averaged number density of the non-thermal electrons. This can be attributed to fixing $\int_{p_1}^{p_2}f_\mathrm{e}(p)\,p^2\,dp=1$ and the resulting normalization of $f_\mathrm{e}(p)$. The corresponding lower limits of the magnetic field strength are estimated from the known synchrotron power, and found to be well within the limits of measurements through Faraday rotation. To estimate the diffuse magnetic field strength, we use the equipartition argument to relate the magnetic field strength with the non-thermal particle density, and assume a radially dependent $B(r)$ to estimate a volume-averaged field strength within a spherical volume of radius $0.5\,$Mpc. The limit on the magnetic field strength increases with a decrease in the electron momentum considered, since the electrons with lower energy require larger magnetic fields to produce the same synchrotron flux density at 1.4 GHz, as seen from the formulation in Section \ref{sec:sync}.

\paragraph{}

The resulting magnetic field profile (Figure \ref{fig:neBprofile}) is shallower than the commonly assumed beta-model profile in the literature, staying within the same order-of-magnitude inside the volume bounded by the $r_{500}$. This has followed from our choice of the GNFW profile to model the electron pressure and subsequently $y_\mathrm{nth}(r)$ [Eqs. (\eqref{eq:pr}, \eqref{eq:yr})], which is steeper than the beta model within the $r_{500}$. We have also assumed isothermality for the electrons (pseudo-temperature for the non-thermal electrons), i.e., the same power-law distribution throughout the cluster volume. These assumptions on the GNFW profile and isothermality do not reflect the dynamic environments within the ICM. Rather, we simply consider them as a reasonable set of assumptions considering the current state of knowledge, and the spatial resolution and sensitivities of the CMB experiments.
 
\paragraph{}

Our forecasts indicate that with upcoming experiments such as SO and FYST, improved constraints on $y_\mathrm{nth}$ and $B$ can be obtained. For the most simplistic single-slope power law models (such as S1 with $p_1 = 30$), the prospective constraints on $y_\mathrm{nth}$ with upcoming data would require a central $B_0$ value $>2\,\mu$G to reconcile with the observed synchrotron power (Table \ref{tab:results2}). This is at the limit of some of the recent measurements of the central magnetic field strength using FRM, e.g., \cite{Govoni2017} infer $B_0=1.5\pm 0.2 \;\mu$G in the GC Abell 194.
\paragraph{}
When we consider lower $p_1$ (or $p_\mathrm{br}$), it results in fewer electrons in the higher-momentum tail of the non-thermal electron distribution which emit synchrotron radiation at 1.4 GHz, thus requiring higher magnetic field strengths when assuming the same synchrotron power. Hence, lowering $p_1$ results in higher $B$ estimates which are in tension with the current data. The lower limits on $B_0$ and $\bar{B}$ estimated as a function of $p_1$ and $p_\mathrm{br}$, assuming \plk and SO+FYST sensitivities, and a fixed synchrotron power at 1.4 GHz, are plotted in Figure \ref{fig:p1-pbr-B}. The assumption of a single power-law with $p_1<30$ for the non-thermal electron momentum distribution results in lower limits of $B_0>3\,\mu$G assuming SO+FYST sensitivities. We can thus state that SO$+$FYST data will be able to rule out some of these simplistic models for the non-thermal particle distribution in GCs. This can prove to be extremely useful in discerning the acceleration mechanisms and physical extent of the non-thermal electron population in GCs within the next few years. 

\paragraph{}

It is worth highlighting that these future constraints, with upcoming CMB survey data, are obtained with the parameters of the same 62 galaxy clusters, in other words, assuming a RH sample of 62 clusters within a similar mass and radio power range. As new observations with LOFAR and other low-frequency surveys are rapidly improving the number of known RHs, both in the lower-mass regimes and at higher redshifts (e.g. \cite{Botteon2022} \cite{DiGennaro2021}), better statistical accuracy will be available with larger RH sample when leveraging the future CMB data for the ntSZ effect. However, more accurate forecasts utilizing a larger cluster sample size would require careful modelling of the scaling of the RH power with both cluster mass and redshift, which we have left out for a future study.

\appendix

\section{Modeling relativistic tSZ and kSZ}
\label{sec:modelsz}

\subsection{Relativistic tSZ}
\label{app:rSZ}
The momentum distribution of scattering electrons which give rise to the $\mathrm{tSZ}_\mathrm{rel}$ effect are modelled using the Maxwell-J\"{u}ttner distribution. Here, a distribution of electron momenta can be described in terms of the normalized thermal energy-parameter, $\Theta = \frac{k_\mathrm{B}T_\mathrm{e}}{m_\mathrm{e} c^2}$, as
\begin{equation}
    f_\mathrm{e,th}(p;\Theta) = \frac{1}{\Theta K_2(1/\Theta)}p^2 \mathrm{exp}(-\frac{\sqrt{1+p^2}}{\Theta}) ,
    \label{eq:maxjutt}
\end{equation}
where $K_v(x)$ denotes the modified Bessel function of the second kind which is introduced for appropriate normalization of the distribution. 
The total IC spectrum for a Planck distribution of photons with  specific intensity of CMB is computed as,
\begin{equation}
    \delta i(x) = \Bigg(\int_0^\infty\int_{-\infty}^\infty f_\mathrm{e,th}(p;\Theta)\: K(e^\mathrm{s};p) e^\mathrm{s} \: i(x/e^\mathrm{s})\: ds\: dp - i(x)\Bigg) I_0\tau_\mathrm{e,th} .
    \label{eq:thermal1}
\end{equation}
Eq. (\ref{eq:thermal1}) is numerically integrated employing the following limits on the integrands,
\begin{equation}
    \delta i(x) = \Bigg(\int_{p_1}^{p_2}\int_{-s_{m}(p_1)}^{s_{m}(p_2)} f_\mathrm{e,th}(p;\Theta) K(e^\mathrm{s};p) e^\mathrm{s}\: \Bigg[\frac{(x/e^\mathrm{s})^3}{(e^{x/e^\mathrm{s}}-1)}\Bigg]\: ds\: dp - \Bigg[\frac{x^3}{(e^x-1)}\Bigg] \Bigg) I_0\frac{m_\mathrm{e} c^2}{k_\mathrm{B}T_\mathrm{e}} y_\mathrm{th} ,
    \label{eq:thermal2}
\end{equation}
where $s_m(p) = 2\mathrm{arcsinh}(p)$, $K(e^s;p)$ is described in Eq. (\ref{eq:ksp}) and we have used $\tau_\mathrm{e,th}=\frac{m_\mathrm{e} c^2}{k_\mathrm{B}T_\mathrm{e}} y_\mathrm{th}$, with $T_\mathrm{e}$ representing the temperature of the scattering electrons.
Eq. (\ref{eq:thermal2}) provides the correct estimation of the tSZ effect with relativistic corrections ($\mathrm{tSZ_{rel}}$) for electron energies $> 1\,$keV, which is the case with the ICM.
\paragraph{} In order to test that the implementation of the photon redistribution function to estimate the SZ effect for a given distribution of electron momenta is correct, $\mathrm{tSZ_{rel}}$ is computed using Eq. (\ref{eq:thermal2}) and compared with the methods provided in \cite{chlubaszpack} and \cite{Itoh1998}.
\subsection{kSZ}
\label{app:ksz}
The kSZ effect is the distortion in the specific intensity or temperature of the CMB due to scattering of the CMB photons by free electrons undergoing bulk motion. The distortion in specific intensity of CMB due to the kSZ effect is written as
\begin{equation}
    \Delta I_\mathrm{kSZ} =  -I_0\, \tau_\mathrm{e}\,\Big(\frac{v_\mathrm{pec}}{c}\Big)\frac{x^4e^x}{(e^x-1)^2},\quad x=\frac{h\nu}{k_\mathrm{B}T_\mathrm{CMB}},
    \label{eq:iksz}
\end{equation}
where $I_0$ is the specific intensity of the CMB, $v_\mathrm{pec}$ is the peculiar velocity associated with the cluster along line-of-sight and $\tau_\mathrm{e}$ is the optical depth due to the free electrons. The optical depth can be expressed in terms of the Compton-y parameter ($y_\mathrm{th}$) as
\begin{equation}
    \tau_\mathrm{e} = \int\sigma_T\,n_\mathrm{e}\,dl = \frac{m_\mathrm{e}c^2}{k_\mathrm{B}T_\mathrm{e}}y_\mathrm{th},
\end{equation}
and a parameter analogous to $y_\mathrm{tSZ}$ for the kSZ effect is defined as
\begin{equation}
    y_\mathrm{kSZ} = \tau_\mathrm{e}\,\Big(\frac{v_\mathrm{pec}}{c}\Big) = \frac{m_\mathrm{e}c^2}{k_\mathrm{B}T_\mathrm{e}}\,\Big(\frac{v_\mathrm{pec}}{c}\Big)y_\mathrm{th}.
    \label{eq:yksz}
\end{equation}

\subsection{Bandpass corrections}
\label{sec:bandpass}
Application of bandpass corrections is necessary to reduce errors due to  systematic effects introduced by the variation in spectral response of each detector in a frequency channel. The spectral response of the detectors was measured through ground-based tests \cite{LFIresponse, PlanckHFIresponse}.  Following the formalism presented in \cite{PlanckHFIresponse}, the bandpass corrected SZ-spectra are computed as
\begin{equation}
    \Delta \tilde{I}_\mathrm{tSZ_{rel}}(x) = y_\mathrm{th}\:I_0\:\frac{\int d\nu\: \tau(\nu)\:g(x,T_\mathrm{e})}{\int d\nu\: \tau(\nu)\: \big(\frac{\nu_c}{\nu}\big)},
    \label{eq:bandpasstsz}
\end{equation}
for the $\mathrm{tSZ_{rel}}$ effect spectrum for scattering electron temperature $T_\mathrm{e}$ and
\begin{equation}
    \Delta \tilde{I}_\mathrm{ntSZ}(x) = y_\mathrm{nth}\:I_0\:\frac{\int d\nu\: \tau(\nu)\:\tilde{g}(x)}{\int d\nu\: \tau(\nu)\: \big(\frac{\nu_c}{\nu}\big)},
    \label{eq:bandpassntsz}
\end{equation}
for the ntSZ effect spectra where $\Delta \tilde{I}_\mathrm{ntSZ}(x)$ is computed separately for each of the non-thermal electron distributions considered in this work. In the equations, $\nu_c$ denotes the central frequency of the frequency bands, and $\tau(\nu)$ is the spectral transmission\footnote{2018 release of filter bandpass transmissions used in this work are available at: \url{https://wiki.cosmos.esa.int/planck-legacy-archive/index.php/The_RIMO}} at frequency $\nu$.

\acknowledgments
We thank the referee for valuable feedback on the manuscript.
We thank E. Komatsu for extensive feedback on the draft. We thank T. En\ss lin, J. Erler, and S. Majumdar for useful discussions. This work made use of \texttt{astropy}\footnote{http://www.astropy.org} \cite{astropy:2013, astropy:2018, astropy:2022}, \texttt{scipy} \cite{scipy}, \texttt{healpy} \cite{healpy} and \texttt{emcee} \cite{emcee}.

\bibliography{main}

\end{document}